\documentclass[twocolumn,english,twocolumn,prl,amssymb,amsmath,showpacs]{revtex4}
\usepackage[latin9]{inputenc}
\usepackage{graphicx,color}
\usepackage{amssymb}
\usepackage[bookmarks]{hyperref}
\usepackage{babel}

\newcommand{\be}{\begin{equation}}
\newcommand{\ee}{\end{equation}}
\newcommand{\ua}{\uparrow}
\newcommand{\da}{\downarrow}
\newcommand{\stag}{\mathcal{M}_{s}}

\begin{document}

\title{Dynamical quantum phase transitions in systems with broken-symmetry phases}

\author{M. Heyl}
\affiliation{Institute for Quantum Optics and Quantum Information of the Austrian Academy of Sciences, 6020 Innsbruck, Austria}
\affiliation{Institute for Theoretical Physics, University of Innsbruck, 6020 Innsbruck, Austria}

\begin{abstract}

In this work it is shown that dynamical quantum phase transitions in Loschmidt echos control the nonequilibrium dynamics of the order parameter after particular quantum quenches in systems with broken-symmetry phases. A direct connection between Loschmidt echos and the order parameter dynamics is established which links nonequilibrium microscopic probabilities to the system's macroscopic dynamical properties. These concepts are illustrated numerically using exact diagonalization for quantum quenches in the XXZ chain with initial Ne\'el states. An outlook is given how to explore these predictions experimentally with ultra-cold gases in optical lattices.

\end{abstract}

\pacs{64.70.Tg,05.30.Rt}

\maketitle


\emph{Introduction:-} In equilibrium thermodynamic phase transitions are accompanied by nonanalyticities in thermodynamic potentials leading to abrupt changes in the macroscopic physical properties.
Recently, broad evidence has been provided for a potential generalization of this fundamental concept to nonequilibrium quantum real-time evolution: the relaxational dynamics of observables can exhibit abrupt changes by varying external control parameters suggesting the possibility of different dynamical phases~\cite{Barankov2006,Yuzbashyan2006,Barmettler2009gd,Eckstein2009wj,Eckstein2009yp,Schiro2010gj,Barmettler2010,Gambassi2011,Schiro2011,Mathey2011nf,Calabrese2012b,Heyl2013a,Sciolla2010jb,Sciolla2011,Igloi2012}. While the observed phenomenology is to a large extend compatible with a dynamical analogue the underlying principles are unclear and a framework allowing to address fundamental questions such as universality is missing. It is the purpose of this work to link these observations to a recently introduced concept of a dynamical quantum phase transition (DQPT)~\cite{Heyl2013a} for systems with broken-symmetry phases thereby opening a path to dynamical criticality on general grounds.

Systems with broken-symmetry phases in equilibrium constitute one subclass of  models in which the possibility of dynamical phase transitions has been suggested~\cite{Barmettler2009gd,Calabrese2012b,Heyl2013a,Sciolla2011,Igloi2012}. This is due to a generic feature observed in the nonequilibrium dynamics whenever the system is initially prepared in the broken-symmetry phase. In consequence of a sudden switching of an external parameter $\lambda$, a so-called quantum quench, beyond a critical value $\lambda_c$ the decay of the equilibrium order parameter has been found to show an abrupt change from monotonic to oscillatory. 

In this work a link between this sharp appearance of the order parameter oscillations and DQPTs is established. In Ref.~\cite{Heyl2013a} it has been shown that the nonequilibrium real-time evolution after a quantum quench can generate nonanalyticities as a function of time in Loschmidt amplitudes
\be
	G(t) = \langle \psi_0 | e^{-iHt} | \psi_0 \rangle,
\label{eq:defLoschmidtAmplitude}
\ee
where $|\psi_0\rangle$ is the initial state (typically the ground state of a Hamiltonian $H_0$ at  $\lambda_0$) and $H$ the Hamiltonian at the final value $\lambda$ of the switched parameter. In the meantime these DQPTs at critical times have been found in a variety of different systems~\cite{Pollmann2010dv,Heyl2013a,Karrasch2013,Fagotti2013,Andraschko2014,Hickey2014,Vajna2014,Kriel2014}. Importantly, it has been shown that these transitions are stable against weak perturbations that preserve the symmetries of the model~\cite{Karrasch2013,Kriel2014}. Notice that dynamical transitions have also been found in different contexts~\cite{Hedges2009kn,Garrahan2010xw,Diehl2010,Mitra2012}.

The discovery of these DQPTs open the possibility to study fundamental questions such as scaling and universality in quantum real-time evolution. Here, a major challenge is to link the microscopic probabilities or amplitudes $G(t)$ that host the DQPTs to macroscopic properties which are the quantities of primary interest from an experimental perspective. Although there is numerical evidence for such a link for particular systems~\cite{Heyl2013a,Calabrese2012b,Vajna2014} the underlying mechanism, however, is still unclear.

It is the aim of this work to develop a theory linking DQPTs to the dynamics of local observables thereby establishing a connection between nonequilibrium microscopic probabilities and macroscopic properties. It will be shown that this link is provided by a dynamical analogue to equilibrium critical regions in the vicinity of quantum critical points thereby further bridging the gap between DQPTs and equilibrium criticality. The main concepts will be illustrated for the XXZ chain, the underlying ideas, however, are far more general and can be applied also to other systems as will be summarized at the end of this letter.


\emph{XXZ chain:-} These concepts will be studied exemplary for anisotropy quenches in the XXZ chain:
\be
	H_\Delta = J \sum_{l=0}^{N-1} \left[ S_l^x S_{l+1}^x + S_l^y S_{l+1}^y + \Delta S_l^z S_{l+1}^z \right],
\label{eq:defXXZModel}
\ee
with $J>0$ antiferromagnetic, $N$ the number of lattice sites, and $S_l^\alpha$, $\alpha=x,y,x$, spin-1/2 operators. In equilibrium this model exhibits a quantum critical point at $\Delta =1$ separating a gapless  ($\Delta <1$) from a gapped phase ($\Delta >1$) with antiferromagnetic order. The order parameter of this transition is the staggered magnetization
\be
	\mathcal{M}_s = \frac{1}{N} \sum_{l=0}^{N-1} (-1)^l S_l^z.
\label{eq:defStaggeredMagnetization}
\ee
Nonequilibrium dynamics will be generated via a quantum quench~\cite{Polkovnikov2011kx}. The system is initialized in a Ne\'el state:
\be
	|\psi_0\rangle = |\ua \da \rangle = |\uparrow \downarrow \uparrow \downarrow \dots \rangle,
\ee
which is equivalent to preparing the system in the ground state of the XXZ chain at initial anisotropy $\Delta_0\to\infty$. The quantum real-time evolution is driven by the final Hamiltonian $H=H_\Delta$ at anisotropy $\Delta<\infty$. The numerical results are obtained using exact diagonalization (ED) based on a Lanczos tridiagonalization of the Hamiltonian with full reorthogonalization~\cite{Cullum2002}. For the numerical calculations periodic boundary conditions have been chosen.

\begin{figure}
\centering
\includegraphics[width = 0.9\columnwidth]{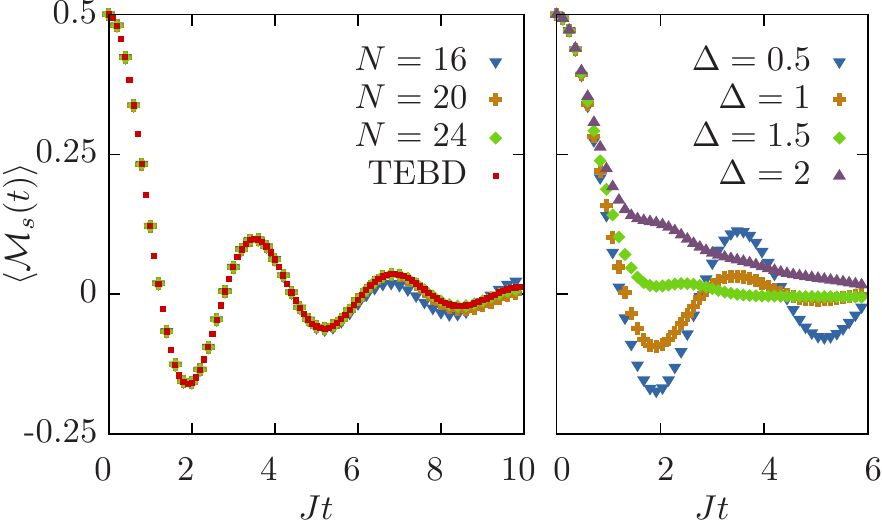}
\caption{(color online) (a) Oscillatory decay of the staggered magnetization in the XXZ chain for initial Ne\'el states and final anisotropies $\Delta=0.6$ obtained using exact diagonalization (ED) for different system sizes $N$. Up to times $Jt=10$ the ED data for $N=24$ matches the thermodynamic limit result from TEBD obtained using the ALPS libraries~\cite{Bauer2011}. (b) Increasing the anisotropy from $\Delta = 0.5$ to $\Delta = 2$ (for $N=24$) the decay of the staggered magnetization changes from oscillatory to monotonic, see also Ref.~\cite{Barmettler2009gd}.}
\label{fig:1}
\end{figure}

For initial Ne\'el states the staggered magnetization shows a transition from a monotonic long-time decay to oscillatory as soon as $\Delta<1$ crosses the equilibrium phase boundary~\cite{Barmettler2009gd,Barmettler2010}. Within some intermediate regime $1<\Delta\lesssim 2$ the long-time behavior is monotonic, on transient time scales, however, oscillatory behavior can be found~\cite{Barmettler2009gd}. In Fig.~\ref{fig:1} ED data illustrates the oscillatory decay for quenches to a final $\Delta=0.6$ and the transition to monotonic decay by increasing the anisotropy. Moreover, analytical and numerical results show that the model also exhibits real-time nonanalyticities in Loschmidt amplitudes and thus DQPTs~\cite{Fagotti2013,Andraschko2014}. 


\emph{Spectral decomposition:-} If the initial Hamiltonian and the order parameter commute, both observables can be measured simultaneously such as in the case of the initial Ne\'el state in the XXZ chain where $[H_{\Delta_0},\stag]=0$ at $\Delta_0\to\infty$. It is therefore possible to decompose the order parameter, e.g., the staggered magnetization $\stag$ for the XXZ chain, spectrally during its dynamical evolution:
\be
	\langle \mathcal{M}_s(t) \rangle = \int d\varepsilon \,\, \mathcal{M}_s(\varepsilon,t) \,\, P (\varepsilon,t).
\label{eq:defSpectralDecomposition}
\ee
Here, $P(\varepsilon,t)$ is the probability distribution that the system has energy density $\varepsilon$ at time $t$ (with energies measured by $H_{\Delta_0}$) and $\mathcal{M}_s(\varepsilon,t)$ is the contribution to the full expectation value $\langle\stag(t)\rangle$ from energy density $\varepsilon$.  The energy density distribution $P(\varepsilon,t)$ is defined by
\be
	P(\varepsilon,t) = \sum_{\nu} |\langle E_\nu |\psi_0(t)\rangle|^2 \delta(E_\nu/N-\varepsilon),
\ee
with $|\psi_0(t)\rangle = e^{-iHt}|\psi_0\rangle$ the time evolved initial state and $|E_\nu\rangle$ a complete set of eigenstates of the initial Hamiltonian $H_{\Delta_0}$ with the respective energies $E_\nu$. For technical details, see below. The zero of energy is chosen such that the ground state of $H_{\Delta_0}$ has vanishing energy.

It is important to emphasize that in the context of Eq.~(\ref{eq:defSpectralDecomposition}) energies are \emph{not} measured with the final Hamiltonian but rather with the initial one. Thereby, an ``exclusive'' perspective~\cite{Campisi2011kx} is chosen in which the  perturbation which generates the dynamics is not included into the system's internal energy. This choice is based on the observation that all properties addressed in this work, the staggered magnetization as the order parameter for the antiferromagnetic phase and the Loschmidt amplitude as a ground state to ground state overlap, are rather connected to the initial than the final Hamiltonian.


\emph{Dynamical phase transitions:-} In the following, it will be shown that $P(\varepsilon \to 0,t)=\mathcal{L}(t)$ is a Loschmidt echo $\mathcal{L}=|G(t)|^2$ and as such inherits the DQPT. Most importantly, dynamical transitions in $P(0,t)$ directly result in real-time nonanalyticities of $\stag(0,t)$. These zero energy transitions in $\stag(0,t)$ although smoothed extend their influence to nonzero energies $\stag(\varepsilon>0,t)$ leading to an oscillatory decay of the full expectation value $\langle \stag(t) \rangle$. This connection directly generalizes to other systems with broken-symmetry phases.

\begin{figure}
\centering
\includegraphics[width = 0.9\columnwidth]{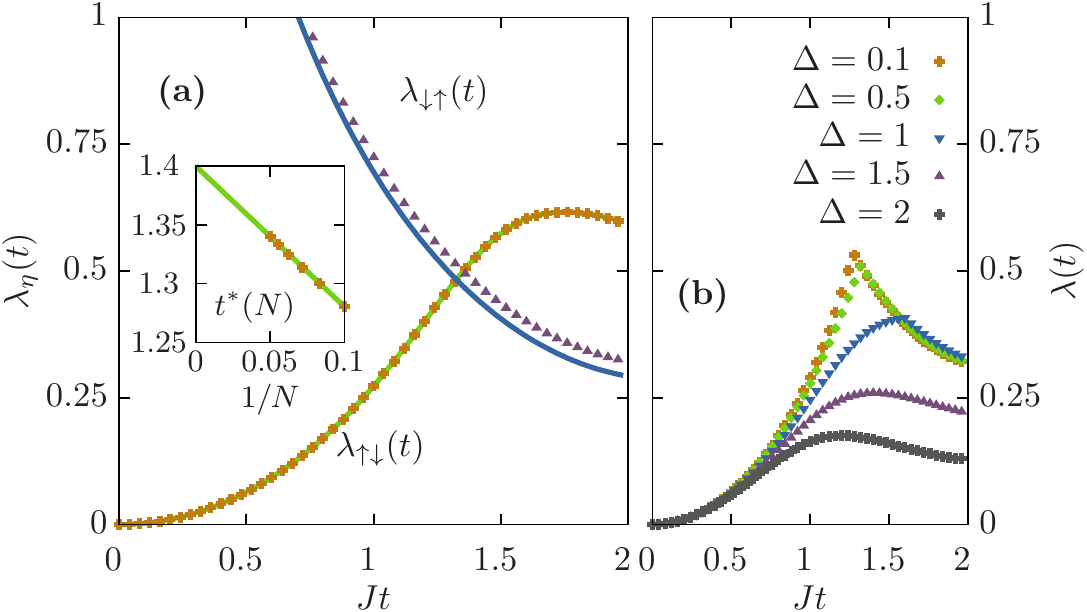}
\caption{(color online) Dynamical quantum phase transition (DQPT) from ED. (a) The two overlap rate functions $\lambda_\eta(t)$ with $\eta=\ua\da,\da\ua$, each for two different system sizes $N=16$ (lines) and $N=24$ (dots) at anisotropy $\Delta = 0.6$. At each $N$ the $\lambda_\eta(t)$ cross each other indicating a DQPT in  $\lambda(t)=\min_\eta \lambda_\eta(t)$. While the $\lambda_{\ua\da}(t)$ component shows no appreciable finite-size scaling, the curve of $\lambda_{\da\ua}(t)$ shifts to larger times for increasing $N$. The location of the DQPT in the thermodynamic limit can be estimated by studying the system size dependence of the intersection point $t^\ast(N)$ of $\lambda_{\ua\da}(t)$ and $\lambda_{\da\ua}(t)$, see inset. A fit to data gives a DQPT at $t^\ast\approx 1.40/J$.  In (b) the behavior of $\lambda(t)$ is shown for different anisotropies $\Delta$ at $N=24$ indicating that for increasing $\Delta$ the DQPT is shifted to larger times eventually moving beyond $Jt=2$. For times $Jt>2$ (not shown) finite-size effects in the overlaps, but not the staggered magnetization, see Fig.~\ref{fig:1}, become substantial preventing a detailed analysis in this regime.}
\label{fig:2}
\end{figure}

Due to the twofold degeneracy of the ground state manifold in $\mathbb{Z}_2$ broken-symmetry phases the zero energy density $\varepsilon\to0$ limit of the energy distribution contains two contributions which in the present XXZ chain are
\be
	P(0,t) = \mathcal{L}_{\uparrow \downarrow}(t) + \mathcal{L}_{\downarrow \uparrow}(t),
\label{eq:P0t}
\ee
with $\mathcal{L}_{\eta}(t) = |\langle \eta |\psi_0(t) \rangle|^2$ and $\eta = \uparrow \downarrow,\downarrow\uparrow $ labeling the two degenerate ground states of $H_{\Delta_0}$. For large systems $N \gg 1$ each of the microscopic probabilities $\mathcal{L}_{\eta}(t)$ obeys a large deviation scaling~\cite{Touchette2009lb} $\mathcal{L}_{\eta}(t)=\exp[-N\lambda_{\eta}(t)]$ with $\lambda_\eta(t)$ intensive~\cite{Silva2008gj,Gambassi2012a,Heyl2013a}. As a consequence, one of the two overlaps will always dominate:
\be
	P(0,t)  = e^{-N\lambda(t)}, \,\, \lambda(t) = \min_{\eta} \lambda_\eta(t),
\label{eq:dptInP}
\ee
up to exponentially small corrections. In Fig.~\ref{fig:2} plots of the rate functions $\lambda_\eta(t)$ are shown at $\Delta=0.6$ for different system sizes $N$. At each $N$ the two rate functions $\lambda_\eta(t)$ cross at a time $t^\ast(N)$ yielding a kink in $\lambda(t)$ due to the sudden switching between the two broken-symmetry sectors. The location of the intersection point in the thermodynamic limit can be found by finite-size scaling which yields $t^\ast \approx 1.40/J$, see Fig.~\ref{fig:2}. In the context of its definition according to Ref.~\cite{Heyl2013a} the system exhibts a \emph{dynamical} quantum phase transition at $t^\ast$. It is important to emphasize that in this way it is possible to detect a DQPT occurring only in the thermodynamic limit~\cite{Heyl2013a} from finite-size ED data with high accuracy.


\emph{Energy-resolved staggered magnetization:-} As $\stag$ and $H_{\Delta_0}$ commute at $\Delta_0\to\infty$ both observables can be measured simultaneously such that
\be
	\langle \stag(t)\rangle = \int d\varepsilon \int dm \,\, m \, P(\varepsilon,m;t),
\label{eq:jointPDF}
\ee
with $P(\varepsilon,m;t)$ the joint distribution function that the system has energy density $\varepsilon$ and staggered magnetization density $m$ at time $t$. Eq.~(\ref{eq:jointPDF}) reflects the potential to perform the following  measurement sequence: first a projective energy measurement onto the eigenstate $|E\rangle$ with energy density $\varepsilon=E/N$ followed by a measurement of the staggered magnetization. 

For $N \gg 1$ the distribution $P(\varepsilon,m;t)$ satisfies a central-limit theorem~\cite{supp} such that at a given $\varepsilon$ only a narrow region (vanishingly small in the thermodynamic limit) contributes dominantly in the vicinity of $m=\stag(\varepsilon,t)$ where $P(\varepsilon,m;t)$ becomes maximal. This yields the desired result in Eq.~(\ref{eq:defSpectralDecomposition}) with the identification $P(\varepsilon,t) = \int dm P(\varepsilon,m;t)$. Using large-deviation theory~\cite{Touchette2009lb} one can compute $\stag(\varepsilon,t)$ as the expectation value $\stag(\varepsilon,t) = \langle \psi_0(t,s) | \stag | \psi_0(t,s)\rangle$ in the state $ |\psi_0(t,s)\rangle = [\mathcal{N}(s,t)]^{-1/2} e^{-H_{\Delta_0}s/2} |\psi_0(t)\rangle$ with $\mathcal{N}(s,t) =\langle \psi_0(t) |e^{-H_{\Delta_0}s}|\psi_0(t)\rangle$ and $s=s(\varepsilon,t)$ given by the solution of the equation $\varepsilon = N^{-1} \langle \psi_0(t,s) | H_{\Delta_0} | \psi_0(t,s)\rangle$~\cite{supp}.
\begin{figure}
\centering
\includegraphics[width = 0.9\columnwidth]{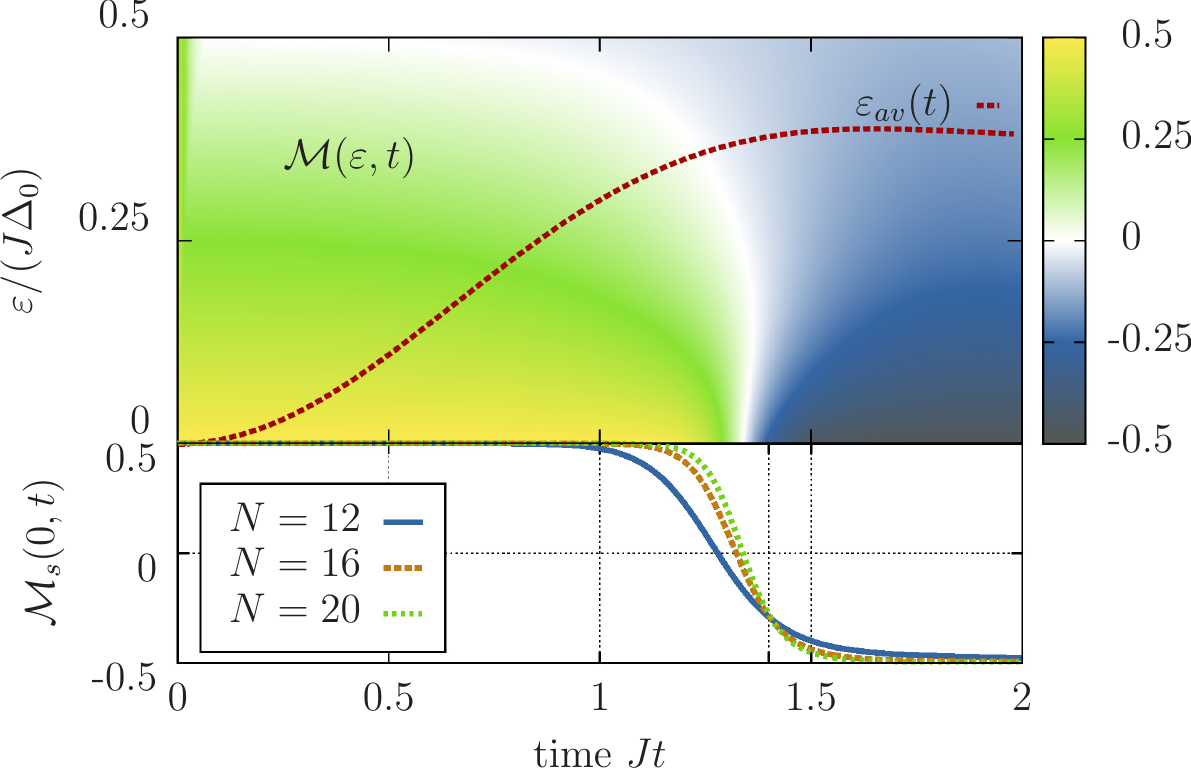}
\caption{(color online) Upper false-color plot: Energy-resolved staggered magnetization $\stag(\varepsilon,t)$ in the $\varepsilon-t$ plane for quenches to a final anisotropy $\Delta = 0.6$ with $N=20$. The DQPT at $\varepsilon=0$ gets smeared at nonzero energies but its influence, a change in sign of the staggered magnetization, extends to $\varepsilon>0$. The dominant contribution to $\langle \stag(t) \rangle$ comes from a narrow interval in the vicinity of $\varepsilon =\varepsilon_\mathrm{av}(t)$, see main text, the dotted line depicts its dynamics. Lower plot: Zero-energy limit $\stag(0,t)$ for different system sizes $N$. For increasing $N$ the change in staggered magnetization becomes sharper eventually yielding a jump as one can directly infer from Fig.~\ref{fig:2}. As in Fig.~\ref{fig:2} one can see that the point $t^\ast(N)$ where $\stag(0,t)$ becomes zero (where $\lambda_{\ua\da}(t)=\lambda_{\da\ua}(t)$) shifts to larger times for larger $N$. A close inspection shows that there is a point $t\approx1.40/J$ in time (indicated by a dashed line) where $\stag(0,t)$ for all $N$ considered intersect each other which yields the location of the DQPT as estimated in Fig.~\ref{fig:2}.}
\label{fig:dcr}
\end{figure}

In Fig.~\ref{fig:dcr} a false-color plot of $\stag(\varepsilon,t)$ obtained via ED is shown in the $\varepsilon-t$ plane. Additionally, a finite-size scaling of the staggered magnetization at zero energy is included revealing for times $t<t^\ast$ that $\stag(0,t)\to1/2$ whereas for $t>t^\ast$ that $\stag(0,t)\to-1/2$. At $t\approx t^*$  there is a crossover which becomes sharper for increasing system sizes. In the thermodynamic limit this yields a jump because from Fig.~(\ref{fig:2}) one can directly infer that at $t=t_\ast$ the dominant contribution in the zero energy sector switches from $\eta=\ua\da$ with staggered magnetization $+1/2$ to $\eta=\da\ua$ with staggered magnetization $-1/2$. Thus, the DQPT in the Loschmidt amplitude directly translates into a real-time nonanalyticity in the zero-energy limit $\stag(0,t)$ of the order parameter.

How does this nonanalyticity at zero energy influence the dynamics of the full expectation value $\langle \stag(t)\rangle$ of the staggered magnetization? In the thermodynamic limit the dominant contribution to $\langle \stag(t)\rangle$ comes from a narrow interval in the vicinity of $\varepsilon = \varepsilon_\mathrm{av}(t)=N^{-1} \langle H_0(t) \rangle$  due to the central limit theorem such that $\langle \stag(t)\rangle \to \stag(\varepsilon_\mathrm{av}(t),t)$ for $N\to \infty$. In order to assess the influence of the DQPT onto $\langle \stag(t)\rangle$ it is therefore necessary to study the link between $\stag(0,t)$ and $\stag(\varepsilon_\mathrm{av}(t),t)$.

As one can see from Fig.~\ref{fig:dcr}, the real-time nonanalyticity gets smeared at nonzero energies. Its influence, however, extends to $\varepsilon>0$ as a matter of continuity: the change in sign of $\stag(\varepsilon>0,t)$ is not abrupt any more, but spans over a time interval of nonzero length. The larger the energy density the larger the region in the $\varepsilon-t$ plane which is controlled by the zero-energy real-time nonanalyticity. This extends up to energy densities $\varepsilon_\mathrm{av}(t)$ demonstrating that DQPTs control the sign change of the order parameter and as a consequence  its oscillatory decay. Notice the strong similarity to critical regions at equilibrium quantum phase transitions by associating energy density with temperature and time with the control parameter.

It is important to emphasize that, although there is an apparent similarity between Fig.~\ref{fig:dcr} and equilibrium critical regions, it is not clear whether universality and scaling apply for the DQPT in the concrete case studied here. On the one hand, the DQPT due to a switching between the two symmetry-broken sectors is reminiscent to first-order ground state phase transitions in consequence of a level crossing. On the other hand, jumps in derivatives of thermodynamic potentials can also appear for continuous phase transitions such as in the specific heat of the superconducting-normal state transition in BCS theory. Adressing these general questions of scaling and universality as well as a potential classification scheme for the DQPTs constitutes an interesting and important further step. This, however, requires some further detailed analysis which is left open for future work.

The results obtained here for the XXZ chain naturally generalize to other models as long as the following two requirements are satisfied: firstly, the initial Hamiltonian has to exhibit a ground-state degeneracy, e.g., a system in a broken-symmetry phase, such that $P(0,t)$ is a sum over the individual probabilities to be in the one of the respective ground states, see Eq.~(\ref{eq:P0t}). Secondly, the initial Hamiltonian has to exhibit one point in parameter space where it commutes with the order parameter allowing for the spectral decomposition in Eq.~(\ref{eq:defSpectralDecomposition}). This includes a wide range of systems such as Ising models at vanishing transverse field, Bose- or fermionic Hubbard models at vanishing tunneling in the charge-density wave limit, regardless of dimensionality. Systems with topological order are also accessible such as the Kitaev chain which is equivalent to a  one-dimensional Ising chain through an exact mapping.

The connection between DQPTs and macroscopic dynamical properties is a priori not limited to the order parameter alone. For any observable whose expectation value differs in the two symmetry-broken ground states DQPTs in Loschmidt echos potentially impose real-time nonanalyticities in the ground-state manifold as for the zero-energy limit of the order parameter, see Fig.~\ref{fig:dcr}.

\emph{Experiments:-} The considered nonequilibrium scenario can be realized in systems of ultra-cold atoms in optical lattices~\cite{Bloch2008lv}. In the hard-core limit a one-dimensional system of bosonic particles can be mapped onto an XXZ chain $H_\mathrm{exp} = J_{xy}\sum_l [S_l^x S_{l+1}^x + S_l^y S_{l+1}^y] + J_z \sum_l S_l^z S_{l+1}^z$ taking into account nearest-neighbor interactions~\cite{Cazalilla2011}. Contrary to the Hamiltonian in Eq.~(\ref{eq:defXXZModel}) the coupling $J_{xy}$ is ferro- instead of antiferromagnetic which can be compensated for by a unitary transformation $U=\exp[i(\pi/2)\sum_{l=0}^{N/2-1} \sigma_{2l}^z]$ mapping $H_\mathrm{exp}$ onto $H$. Importantly, both the initial state and the observables under study are invariant under $U$ such that the dynamics by $H_\mathrm{exp}$ and $H$ are identical. The initial Ne\'el state corresponds to a characteristic pattern of particles where even sites are occupied by one boson and odd sites are empty. These states can be generated experimentally with high accuracy~\cite{Foelling2007}. The staggered magnetization can be measured via the bosonic density using quantum gas microscopy~\cite{Bakr2009,Sherson2010zo}. For each  experimental image obtained by the quantum gas microscope one can determine the staggered magnetization as well as the energy corresponding to the initial Hamiltonian such that one can build up the full energy-resolved $\stag(\varepsilon,t)$ successively. Loschmidt echos can be obtained experimentally using a recently proposed measurement scheme~\cite{Daley2012,Pichler2013}. 

Although the spectral decomposition in Eq.~(\ref{eq:defSpectralDecomposition}) requires fine-tuning of the system, it will now be argued that the consequences of a nonideal experimental implementation are, in principle, controllable. As already emphasized in the introduction, weak perturbations to the final Hamiltonian don't influence the DQPTs qualitatively~\cite{Karrasch2013,Kriel2014}. Nonzero-temperature effects can be eliminated using post-selection~\cite{Fukuhara2012}. Although for initial states perturbed by weak initial $J_{xy}>0$ the dynamics does not change qualitatively~\cite{Barmettler2010}, order parameter and initial Hamiltonian do not commute. From a single image of the quantum gas microscope, however, one can still compute the energy of this single experimental realization for the ideal initial XXZ chain at $J_{xy}=0$ that commutes with the order parameter. In this way one can measure $\stag(\varepsilon,t)$ as in the ideal case and the errors made are reduced to the initial state preparation solely, but not the initial Hamiltonian itself. 


\emph{Conclusions:-} In this work it has been shown that dynamical quantum phase transitions in Loschmidt echos are directly connected to the order parameter dynamics in systems with broken-symmetry phases. Thereby, a link is established between microscopic probabilities and macroscopic dynamical properties. These concepts have been illustrated using exact diagonalization for the XXZ chain for initial Ne\'el states, but generalize also to other observables and other systems. A potential implementation in systems of ultracold atoms has been outlined that allows to explore the predictions experimentally.


\begin{acknowledgments}
Valuable discussions with Philipp Hauke, Hannes Pichler, Anatoli Polkovnikov, and Stefan Kehrein are gratefully acknowledged. This work has been supported by the Deutsche Akademie der Naturforscher Leopoldina via the grant LPDS 2013-07 and by the Austrian Science Fund FWF (SFB FOQUS F4016). The ED algorithm uses the Armadillo linear algebra libraries~\cite{Sanderson2010}.
\end{acknowledgments}


\bibliographystyle{apsrev}
\bibliography{literatureDQPTXXZ}


\appendix

\section{Spectral decomposition of the staggered magnetization and large deviation theory}
\label{sec:app}

The aim of this appendix is to outline the calculation of the joint probability distribution function $P(\varepsilon,m;t)$ that the system at time $t$ has energy density $\varepsilon$ and staggered magnetization $m$ using large deviation theory~\cite{Touchette2009lb}. This then directly leads to a computational scheme for the calculation of the energy-resolved staggered magnetization $\stag(\varepsilon,t)$.

The joint probability distribution $P(\varepsilon,m;t)$ is defined as:
\be
	P(\varepsilon,m;t)  = \sum_\nu |\langle E_v |\psi_0(t)\rangle|^2 \delta(\varepsilon-e_\nu)\delta(m_{\nu} - m)
\ee
where $e_\nu = E_\nu/N$ is the energy density and $m_{\nu}$ is the staggered magnetization of the state $|E_\nu\rangle$: $H_{\Delta_0}|E_\nu\rangle = N e_\nu |E_\nu\rangle$ and $\stag|E_\nu\rangle = m_\nu |E_\nu\rangle$. The function
\begin{align}
	G(s,\mu;t) = \sum_\nu |\langle E_v |\psi_0(t)\rangle|^2 e^{-s E_\nu} e^{-\mu N m_\nu} \nonumber \\
	= \langle \psi_0(t) | e^{-s H_{\Delta_0}} e^{-\mu N\stag} |\psi_0(t)\rangle
\end{align}
is related to $P(\varepsilon,m;t)$ via
\be
	G(s,\mu;t) = \int d\varepsilon \int dm P(\varepsilon,m;t) e^{-N\varepsilon s} e^{-N m \mu}.
\ee
The generating function $G(s,\mu;t)$ obeys a large-deviation scaling~\cite{Heyl2013a}:
\be
	G = e^{N g(s,\mu;t)}
\ee
with $g(s,\mu;t)$ an intensive function independent of system size $N$. As a consequence of this large deviation scaling the joint distribution function has to be of the following structure~\cite{Touchette2009lb}
\be
	P(\varepsilon,m;t) = e^{-N\theta(\varepsilon,m;t)}
\ee
with the rate function $\theta(\varepsilon,m;t)$ again intensive and given by the following series of Legendre transforms:
\begin{align}
	\theta(\varepsilon,m;t) = -\inf_{\mu} \left[ \mu m + \varphi(\varepsilon,\mu;t) \right], \nonumber \\
	\varphi(\varepsilon,\mu;t) = \inf_{s} \left[ \varepsilon s + g(s,\mu;t) \right].
\end{align}
The corresponding back transformations read:
\begin{align}
	\varphi(\varepsilon,\mu;t) = -\inf_{m} \left[ \mu m +\theta(\varepsilon,m;t) \right] ,\nonumber \\
	g(s,\mu;t) = -\inf_{\varepsilon} \left[ \varepsilon s - \varphi(\varepsilon,\mu;t) \right].
\end{align}
As in thermodynamics the Legendre transform of $\theta(\varepsilon,m;t)$, for example, can be calculated by:
\be
	\varphi(\varepsilon,\mu;t) = - \mu m(\varepsilon,\mu;t) - \theta(\varepsilon, m(\varepsilon,\mu;t) ;t)
\ee
with $m(\varepsilon,\mu;t)$ solving the ``equation of state'':
\be
	 \mu = - \frac{d\theta(\varepsilon,m;t)}{dm}.
\ee
Provided $g(s,\mu;t)$ is differentiable with respect to $s$ and $\mu$ it can be shown that $P(\varepsilon,m;t)$ obeys a central-limit theorem~\cite{Touchette2009lb} which is used in the main text to derive the energy-resolved observables. In order to obtain the energy-resolved staggered magnetization $\stag(\varepsilon,t)$ it is necessary to determine the infimum of $\theta(\varepsilon,m;t)$ over all $m$ which according to the previous formulas is
\be
	\varphi(\varepsilon,0;t) = -\inf_m \left[ \theta(\varepsilon,m;t)\right].
\ee
Thus, the infimum happens at $\mu=0$ such that from the equation of state
\begin{align}
	m^\ast(\varepsilon,t) =- \left. \frac{\partial \varphi(\varepsilon,\mu;t)}{\partial\mu} \right|_{\mu=0} = -\left. \frac{\partial g}{\partial\mu} \right|_{s=s(\varepsilon,\mu;t),\mu=0} \nonumber \\
	= \left. \langle \psi_0(t,s) | \stag(t) |\psi_0(t,s) \rangle\right|_{s=s(\varepsilon,\mu=0;t)}  ,
\end{align}
with
\begin{align}
	|\psi_0(t,s)\rangle = \frac{e^{-s H_{\Delta_0}/2}}{\sqrt{\mathcal{N}}} |\psi_0(t)\rangle,\nonumber \\
	\mathcal{N} = \langle \psi_0(t) | e^{-sH_{\Delta_0}} |\psi_0(t) \rangle,
\end{align}
and $s=s(\varepsilon,\mu=0,t)$ solves the equation of state
\be
	\varepsilon = \frac{1}{N} \langle \psi_0(t,s) | H_{\Delta_0} | \psi_0(t,s) \rangle.
\ee
This yields the formulas presented in the main text.

\end{document}